# Single and double ultrashort laser pulse scattering by spheroidal metallic nanoparticles

P.M. Tomchuk[a], D.V. Butenko[a,b,*]

[a] *Institute for Physics, National Academy of Sciences of Ukraine, 46, Nauky Avenue, Kiev-28, 03680, Ukraine*
[b] *National University of Kyiv-Mohyla Academy, 04070 Kyiv, Ukraine*

**Abstract**

The theory of the ultrashort laser pulse scattered by metallic nanoparticles in the region of surface plasmon resonances is developed in the framework of kinetic approach. For the spheroidal particles, the dependence of the light scattering cross-section on the shape of the particles, the carrier wave frequency, a pulse duration, and other factors is studied. Additionally, an interaction of small metallic particles with double ultrashort pulse is considered. In this case, the energy scattered by the particles demonstrates oscillating behavior when the time delay between pulses changes. Special attention is paid to the contribution of the particle's surface when the particle's size is close to the length of free electron path. The difference between the kinetic approach and Mie theory for non-spherical particles has been shown.

*Keywords:* ultrashort pulse, plasmon resonance, nanospheroid, light scattering, kinetic approach

## 1. Introduction

An enhancement of local electric fields near metallic nanoparticles makes an investigation of these objects topical both for the theory of condensed matter and its various applications. Collective electron effects, observed in the nanoparticles of different geometries, make possible to use them for producing biomarkers and biosensors, optical antennas, subwavelength devices, metamaterials with negative refractive index, integrated circuits, and so on (see, e.g.,[1]-[8]).

Modern optical methods permit to study experimentally both ensembles of nanoparticles and separate nanoobjects. This opens great prospects for research on many-electron dynamics in low-dimensional systems. Observation and investigation of fast processes' dynamics not only in single particles, but also in individual molecules and atoms, becomes possible due to the last achievements

*Corresponding author
*Email addresses:* ptomchuk@iop.kiev.ua (P.M. Tomchuk), daniil.butenko@gmail.com (D.V. Butenko)

in femto- and attosecond optics (see, for example, monograph [9] and literature cited there). A broad class of phenomena can be investigated with the help of ultrashort laser pulses: from nonlinear processes, phase transitions in solids, and the processes of formation and breakdown of chemical bonds to magnetic properties of solids, and extreme-ultraviolet light generation in plasmonic nanostructures [10]-[14].

When the light frequency is close to the frequency of surface plasmon resonances (SPR) light absorption and scattering in metallic nanoparticles extremely enhance. In ellipsoidal nanoparticles there are three such resonances, while particles with spherical shape have only one resonance. Thus, an interaction of light with ellipsoidal nanoparticles strongly depends on their shape. In the case of monochromatic electromagnetic wave, contribution of the SPR to the interaction and absorption sharply increases when the wave frequency is close to one of the frequencies of the SPR. The situation drastically changes when the nanoparticle is irradiated by an ultrashort laser pulse. Since such wave includes many harmonics, it can excite several plasmon resonances simultaneously; and the shorter the pulse, the greater is set of its harmonics.

Earlier we investigated the nanoparticles' shape effect on both absorption and scattering of the monochromatic wave [15, 16]. The peculiar features of the absorption of ultrashort pulses by the metallic nanoparticles were studied. In particular, it was shown that if the size of a non-spherical nanoparticle is smaller than the free electron path, the optical conductivity becomes a tensor, and its components define half-widths of the plasmon resonances [17]. So, accounting of the input of nanoparticles' asymmetry to half-widths of the SPR is not limited only to the influence of depolarization factors.

In this paper we study the effects of metal nanoparticles' shape and size on the scattering of ultrashort laser pulses with carrier frequencies close to frequencies of the SPRs. The main distinction from the previous investigations is considering pairs of laser pulses with equal duration. First pulse excites the plasmon oscillations, and the second one can either enhance or weaken the scattering effect depending on the phase of plasmon oscillations. In this paper we find: (i) cross-section of light scattering by ellipsoidal metallic nanoparticles both for single ultrashort laser pulse and for pair of such pulses; (ii) its dependence on the nanoparticles' shape and size, frequency of a carrier wave, pulse duration and time delay between pulses.

The structure of the paper is the following. In Sec.II we describe the problem and method for its solving. Then in Sec.III, we find components of conductivity tensor by solving a kinetic equation. Finally, Sec.IV contains results and its discussion.

## 2. Formulation of the problem

Let us consider the general case of the ellipsoidal metallic nanoparticle (MN) with three different semiaxes $(R_x, R_y, R_z)$ in the field of ultrashort electromagnetic pulse. Such shape of the particle is convenient because the obtained results can be applied to a broad class of MNs: from disk-shaped to needle-like ones. The electric field of an incident electromagnetic wave is given as:

$$\mathbf{E}(\mathbf{r},t) = \mathbf{E}_0 \exp\left\{-\Gamma^2\left(t - \frac{\mathbf{k}_0\mathbf{r}}{\omega_0}\right)^2\right\}\cos(\omega_0 t - \mathbf{k}_0\mathbf{r}), \tag{1}$$

where $\mathbf{E}_0$ is an amplitude of an electric field, $\omega_0$ and $\mathbf{k}_0$ are the frequency and the wave vector of the carrier wave correspondingly, $\Gamma$ is a quantity proportional to $1/\tau$, where $\tau$ is the pulse duration (an exact relation between $\Gamma$ and $\tau$ will be fixed later), and $\mathbf{r}$ and $t$ describe spatial coordinates and time.

Performing the Fourier transform of (1), we can obtain:

$$\mathbf{E}(\mathbf{r},\omega) = \int\limits_{-\infty}^{\infty} dt\mathbf{E}(\mathbf{r},t)e^{i\omega t} = \frac{\sqrt{\pi}}{2\Gamma}\mathbf{E}_0\left[e^{-(\omega-\omega_0)^2/4\Gamma^2} + e^{-(\omega+\omega_0)^2/4\Gamma^2}\right]e^{i\mathbf{k}_0\mathbf{r}\frac{\omega}{\omega_0}}. \tag{2}$$

The electric field (1) causes an electrical dipole moment inside the particle, and this moment defines scattering processes. It is necessary to remark that we don't take into account the vortex electrical field induced by a magnetic component of the external electromagnetic wave. This field is related to the magnetic scattering (as well as absorption). At the frequencies of plasmon resonance, this mechanism weakly influences the scattering (or absorption).

An intensity of the scattered radiation, induced by the electric moment of the particle, is defined as [18]

$$I_S = \frac{c}{4\pi}\left|[\mathbf{E}' \times \mathbf{H}']\right| R_0^2 d\Omega, \tag{3}$$

where $\mathbf{E}'$ and $\mathbf{H}'$ are electrical and magnetic components of the scattered wave correspondingly, $c$ is a velocity of the light, $R_0$ is a distance from the nanoparticle to the observation point, and $d\Omega$ is a solid angle. An intensity of the scattered radiation should be averaged over the period in the case of a monochromatic wave. For ultrashort laser pulse we will use the full value of intensity:

$$\delta I_S = \int\limits_{-t'/2}^{t'/2} dt I_S(t), \tag{4}$$

where $t'$ is a lifetime of the dipole. Since an intensity $I_S(t) \to 0$ beyond the limits of ultrashort pulse, we can extend integration in (4) from $-\infty$ to $\infty$. Then passing to the Fourier transform of (4) and using (3), we obtain:

$$\delta I_S = \frac{c}{4\pi}R_0^2 d\Omega \int\limits_{-\infty}^{\infty}\frac{d\omega}{2\pi}\left|\left[\mathbf{E}'(R_0,\omega) \times \left(\mathbf{H}'(R_0,\omega)\right)^*\right]\right|. \tag{5}$$

Thus, we need to know electric and magnetic vectors $\mathbf{E}'$, $\mathbf{H}'$. As was stated above, the electromagnetic field, scattered by the particle, is defined by its dipole moment (at the distances much larger than a wavelength). We can use a relation between the density of high-frequency current inside the metallic particle and electrical dipole moment [18] to define the last one:

$$\frac{\partial}{\partial t}\mathbf{d}(t) = \int\limits_V d^3r'\mathbf{j}(\mathbf{r}', t). \tag{6}$$

Here $\mathbf{d}$ is an electrical dipole moment of metallic particle with volume $V$, $\mathbf{j}$ is the density of high-frequency current.

A high-frequency current inside the particle is caused by an internal electric field $\mathbf{E}_{in}$, induced by the incident electromagnetic field of the laser pulse (1). This internal field $\mathbf{E}_{in}$ is spatially uniform if we assume such inequality:

$$k_0 R_j \ll 1, \tag{7}$$

where $R_j = \max(R_x, R_y, R_z)$. It means that the wavelength $\lambda_0$ of the carrier wave is far above the particle size. Then, taking into account (2), an expression for an internal field is of the form:

$$[E_{in}(\omega)]_j = \frac{[E(0,\omega)]_j}{1 + L_j[\epsilon_{jj}(\omega) - 1]}, \tag{8}$$

where $L_j$ are depolarization factors in the $j$th direction (in the principal axes of an ellipsoid), and $\epsilon_{jj}(\omega)$ is a diagonal component of the dielectric permeability tensor.

## 3. Kinetic equation and the conductivity tensor

Now, to calculate the density of a high-frequency current one should find the electron velocity distribution function. More precisely, it is necessary to find a non-equilibrium addition to the Fermi distribution function, determined by the local field (8). In the linear approach the distribution function can be written as

$$f(\mathbf{r}, \mathbf{v}, t) = f_0(\varepsilon) + f_1(\mathbf{r}, \mathbf{v})e^{-i\omega t}, \tag{9}$$

where $f_0(\varepsilon)$ is the Fermi distribution function. The function $f_1(\mathbf{r}, \mathbf{v})$ can be found as a solution of linearized Boltzmann equation

$$(\nu - i\omega)f_1(\mathbf{r}, \mathbf{v}) + \mathbf{v}\frac{\partial f_1(\mathbf{r}, \mathbf{v})}{\partial \mathbf{r}} + e\mathbf{E}_{in}\mathbf{v}\frac{\partial f_0}{\partial \varepsilon} = 0. \tag{10}$$

The function $f_1(\mathbf{r}, \mathbf{v})$ ought to satisfy the boundary conditions as well. We will take, as is usually done, the boundary conditions that correspond to the diffusive character of scattering

$$f_1(\mathbf{r}, \mathbf{v})|_S = 0,\ v_n < 0, \tag{11}$$

where $v_n$ is the velocity component normal to the particle surface.

The boundary conditions (11) in the case of the ellipsoidal particle depend on angles, and it complicates the evaluation of the (10). It is rather easy to solve (10) and to satisfy the boundary conditions (11) if one passes to the transformed coordinate system, where an ellipsoid with semiaxes $R_x, R_y, R_z$ transforms into a sphere of radius $R$ with the same volume $V$:

$$x_i' = \gamma_i x_i, \; v_i' = \gamma_i v_i, \; \gamma_i = \frac{R}{R_i}, \; R = (R_x R_y R_z)^{1/3}. \tag{12}$$

(Here instead of $x, y, z$ we put $x_i$, where $i = 1, 2, 3$, and the same for velocities).

In the transformed coordinate system eqaution (10) and the boundary conditions (11) can be written as

$$(\nu - i\omega) f_1(\mathbf{r}', \mathbf{v}') + \mathbf{v}' \frac{\partial f_1(\mathbf{r}', \mathbf{v}')}{\partial \mathbf{r}'} + e\mathbf{E}_{in}\mathbf{v}' \frac{\partial f_0}{\partial \varepsilon} = 0, \tag{13}$$

$$f_1(\mathbf{r}', \mathbf{v}')|_{r'=R} = 0, \;\text{ with }\; \mathbf{r}'\mathbf{v}' < 0. \tag{14}$$

The case $\mathbf{r}'\mathbf{v}' < 0$ corresponds to the motion of electrons from the outer surface of the particle (the origin of coordinates is chosen to be at the center of the particle).

Using method of the characteristics, one can easily find a solution of (13) with the boundary conditions (14):

$$f_1(\mathbf{r}', \mathbf{v}') = -e\mathbf{E}_{in}\mathbf{v}' \frac{\partial f_0}{\partial \varepsilon} \frac{1 - \exp[-(\nu - i\omega) t_0(\mathbf{r}', \mathbf{v}')]}{\nu - i\omega}, \tag{15}$$

where the characteristics $t_0(\mathbf{r}', \mathbf{v}')$ can be presented as

$$t_0(\mathbf{r}', \mathbf{v}') = \frac{1}{v'^2} \left[ \mathbf{r}'\mathbf{v}' + \sqrt{(\mathbf{R}^2 - \mathbf{r}'^2)\mathbf{v}'^2 + (\mathbf{r}'\mathbf{v}')^2} \right]. \tag{16}$$

Now, using the solution (15), one can calculate the density of a high-frequency current induced by the electromagnetic wave inside the metallic nanoparticle. The current density is defined by the expression

$$\mathbf{j}(\mathbf{r}, \omega) = 2e \left( \frac{m}{2\pi\hbar} \right)^3 \iiint \mathbf{v} f_1(\mathbf{r}, \mathbf{v}) d^3 v. \tag{17}$$

Starting from now, we consider a special case of the spheroidal nanoparticle (the special case of ellipsoidal one), when $R_x = R_y = R_\perp, R_z = R_\parallel$. Actually we can provide all further integrations in elementary functions for such geometry. By substituting (17) into expression (6), we obtain an electrical dipole moment of a spheroidal nanoparticle [16]:

$$\mathbf{d}(\omega) = \alpha_\perp(\omega)\mathbf{E}(0, \omega) + [\alpha_\perp(\omega) - \alpha_\parallel(\omega)](\mathbf{q}_0 \cdot \mathbf{E}(0, \omega))\mathbf{q}_0, \tag{18}$$

where $\alpha_{\parallel,\perp}(\omega)$ are components of the polarization tensor along and perpendicular to the spheroid's symmetry axis respectively ($\alpha_x = \alpha_y = \alpha_\perp, \alpha_z = \alpha_\parallel$), and $\mathbf{q}_0$ is a unit vector along the spheroid's symmetry axis. The polarization tensor and the tensor of permeability are related via the well-known formula:

$$\alpha_{\perp,\parallel}(\omega) = \frac{V}{4\pi} \frac{\epsilon_{\perp,\parallel}(\omega) - 1}{1 + L_{\perp,\parallel}[\epsilon_{\perp,\parallel}(\omega) - 1]}. \tag{19}$$

The tensor of permeability, in turn, is connected with a complex conductivity tensor through the expression:

$$\epsilon_{\perp,\parallel}(\omega) = 1 - \frac{\omega_{pl}^2}{\omega^2} + i\frac{4\pi}{\omega}\sigma_{\perp,\parallel}(\omega). \tag{20}$$

Here $\omega_{p\,l}$ is the frequency of plasma electrons oscillations in the metal:

$$\omega_{p\,l} = \sqrt{\frac{4\pi n e^2}{m}}, \tag{21}$$

where $e$ and $m$ are the electron charge and mass, respectively, and $n$ is the electron concentration.

As was stated above, when the size of a non-spherical nanoparticle is smaller than the free electron path, the optical conductivity becomes a tensor. Here we consider just such a point. The general formula for the conductivity tensor is given and analyzed in the paper [19]. Here we are interested in the expression for the conductivity tensor in a high-frequency region, as we consider a visible-light range. In this range, components of the conductivity tensor are given by such expressions [17]:

$$\sigma_{\perp}(\omega) = \frac{9}{8}\frac{ne^2 v_F}{mR_{\perp}\omega^2}\varphi_{\perp}, \tag{22}$$

$$\sigma_{\parallel}(\omega) = \frac{9}{8}\frac{ne^2 v_F}{mR_{\perp}\omega^2}\varphi_{\parallel}, \tag{23}$$

where $v_F$ is a Fermi velocity, and such notation is introduced:

$$\varphi_{\perp}(\omega) = \frac{1}{2}\begin{cases} \frac{1}{2}\left(1 + \frac{1}{2e_p^2}\right)\sqrt{1 - e_p^2} + \frac{1}{e_p}\left(1 - \frac{1}{4e_p^2}\right)\arcsin e_p, & \text{if } R_{\perp} < R_{\parallel} \\ \frac{1}{2}\left(1 - \frac{1}{2e_p^2}\right)\sqrt{1 + e_p^2} + \frac{1}{e_p}\left(1+\frac{1}{4e_p^2}\right)\ln\left(\sqrt{1+e_p^2}+e_p\right); & \text{if } R_{\perp} > R_{\parallel} \end{cases} \tag{24}$$

$$\varphi_{\parallel}(\omega) = \frac{1}{2}\begin{cases} \left(1-\frac{1}{2e_p^2}\right)\sqrt{1 - e_p^2}+\frac{1}{2e_p^3}\arcsin e_p, & \text{if } R_{\perp} < R_{\parallel} \\ \left(1 + \frac{1}{2e_p^2}\right)\sqrt{1 + e_p^2} - \frac{1}{2e_p^3}\ln\left(\sqrt{1 + e_p^2} + e_p\right). & \text{if } R_{\perp} > R_{\parallel} \end{cases} \tag{25}$$

where $e_p$ is an eccentricity of the spheroid:

$$e_p^2 = \left|1 - \frac{R_{\perp}^2}{R_{\parallel}^2}\right|. \tag{26}$$

## 4. Scattered field: results and discussion

Thus, we defined all the quantities that characterized a dipole moment, generated by the laser pulse inside the metallic nanoparticle. According to the formula (5), we should know a relationship between a dipole moment and scattered field for calculating an intensity of scattered radiation. In the far-field this relationship can be given by a well-known formula [18]:

$$\mathbf{H}'(R_0,\omega) = \frac{\omega^2}{c^2 R_0}[\mathbf{n}_0 \times \mathbf{d}(\omega)], \tag{27}$$

where $\mathbf{n}_0$ is a unit vector that defines a direction of observation. Besides, in the far-field a scattered wave is transversal, so there is a relationship between electric and magnetic components:

$$\mathbf{E}'(R_0,\omega) = \left[\mathbf{H}'(R_0,\omega) \times \mathbf{n}_0\right]. \tag{28}$$

Using (27), (28), and (5) we obtain an expression for an intensity of scattered radiation:

$$\delta I_S = \frac{cR_0^2}{8\pi^2} d\Omega \int_{-\infty}^{\infty} d\omega \left|\mathbf{H}'(R_0,\omega)\right|^2 = \frac{d\Omega}{8\pi^2 c^3} \int_{-\infty}^{\infty} d\omega \omega^4 \left|[\mathbf{n}_0 \times \mathbf{d}(\omega)]\right|^2. \tag{29}$$

According to (1) $\mathbf{E}(\mathbf{r},t)$ is a real quantity. It means that $\mathbf{E}(\mathbf{r},-\omega) = \mathbf{E}^*(\mathbf{r},\omega)$, and, as a result, similar relationships are fulfilled for $\mathbf{E}'(\mathbf{r},\omega)$, $\mathbf{H}'(\mathbf{r},\omega)$, $\alpha_{\perp,\parallel}(\omega)$, and $\mathbf{d}(\omega)$. Thus, (29) can be rewritten as an integral from 0 to $\infty$:

$$\delta I_S = \frac{d\Omega}{4\pi^2 c^3} \int_0^{\infty} d\omega \omega^4 \left|[\mathbf{n}_0 \times \mathbf{d}(\omega)]\right|^2. \tag{30}$$

Apart from the vector $\mathbf{n}_0$, there are two more unit vectors that define an angle dependence of the intensity of the scattered radiation: $\mathbf{p}_0$ is a unit vector that determines a polarization of the incident wave, and $\mathbf{q}_0$ is a unit vector that defines spheroid's symmetry axis' orientation. According to (18) we obtain a result for a cross-product $[\mathbf{n}_0 \times \mathbf{d}(\omega)]$:

$$\left|[\mathbf{n}_0 \times \mathbf{d}(\omega)]\right|^2 = \left|\alpha_\perp(\omega)[\mathbf{n}_0 \times \mathbf{p}_0] + \left(\alpha_\parallel(\omega) - \alpha_\perp(\omega)\right)(\mathbf{p}_0 \cdot \mathbf{q}_0)[\mathbf{n}_0 \times \mathbf{q}_0]\right|^2 \left|\mathbf{E}(0,\omega)\right|^2 \tag{31}$$

This expression coincides with a similar expression for a monochromatic wave[16]. Besides, it is obviously reduced to the relation for the spherical particle if the components of the polarization tensor $\alpha_\perp$ and $\alpha_\parallel$ become equal ($\alpha_\perp = \alpha_\parallel = \alpha$).

When we deal with a single nanoparticle, its interaction with the laser pulse strongly depends on the wave's polarization (in other words on the angle between $\mathbf{p}_0$ and $\mathbf{q}_0$). The study of this dependence can be found in the paper [17]. Here we consider a collection of chaotically oriented spheroidal nanoparticles. Thus,

(31) ought to be averaged over the directions of vector $\mathbf{q}_0$. After some algebra we obtain an averaged expression:

$$\begin{gathered}\left\langle|[\mathbf{n}_0\times\mathbf{d}(\omega)]|^2\right\rangle_{\mathbf{q}_0}=\frac{1}{15}\left\{2\cdot|\alpha_\perp(\omega)-\alpha_\parallel(\omega)|^2+\frac{1}{2}\left[3\cdot|2\alpha_\perp(\omega)+\alpha_\parallel(\omega)|^2\right.\right.\\ \left.\left.+2\cdot|\alpha_\perp(\omega)|^2+|\alpha_\parallel(\omega)|^2\right]\sin^2\psi\right\}|\mathbf{E}(0,\omega)|^2,\end{gathered}\tag{32}$$

where $\psi$ is an angle between vectors $\mathbf{n}_0$ and $\mathbf{p}_0$. To avoid misunderstanding we should admit that (32) is not a new result, it can be found in a number of sources [20, 21], or it can be obtained from the general equation for light scattered by anisotropic dipole scatterers with random orientations [22].

*4.1. Single ultrashort pulse*

Considering (30) one can obtain the scattering cross-section dividing the intensity of radiation (30) by the energy flux density of an incident wave $\Delta I$:

$$\langle d\Sigma\rangle_{\mathbf{q}_0}=\frac{\langle\delta I_S\rangle_{\mathbf{q}_0}}{\Delta I}=\frac{\frac{d\Omega}{4\pi^2c^3}\int\limits_0^\infty d\omega\omega^4\left\langle|[\mathbf{n}_0\times\mathbf{d}(\omega)]|^2\right\rangle_{\mathbf{q}_0}}{\frac{c}{4\pi}\int\limits_{-\infty}^{\infty}dt\,\mathbf{E}(\mathbf{r},t)\cdot\mathbf{H}(\mathbf{r},t)}.\tag{33}$$

The energy flux density of an incident wave $\Delta I$ can be easily calculated. According to (33) and taking into account (2), we obtain following expression:

$$\Delta I=\frac{c}{8\sqrt{2\pi}\Gamma}\left(1+e^{-\omega_0^2/2\Gamma^2}\right)\mathbf{E}_0^2.\tag{34}$$

Here we need to dwell on the question about the pulse duration (we designate this value as $\tau$). We can define this time using expression (34) in a following way. If we consider the situation when $\omega_0/\Gamma\gg1$, the energy flux density of an incident wave $\Delta I$ divided by the pulse duration $\tau$ ought to be equal to the energy flux density of an incident monochromatic wave divided by its period. Thus, we have a relationship:

$$\frac{\Delta I}{\tau}=\frac{c\mathbf{E}_0^2}{8\pi}\Rightarrow\tau=\sqrt{\frac{\pi}{2}}\frac{1}{\Gamma}\tag{35}$$

If one proceeds to the limit $\Gamma\to0$, an expression for the light scattering cross-section (33) should reduce to the similar expression for a monochromatic wave. For this purpose, we use the next representation for $\delta$ function [23]:

$$\delta(x)=\frac{1}{\sqrt{\pi}}\lim_{\alpha\to0}\frac{1}{\alpha}e^{-x^2/\alpha^2}.\tag{36}$$

And after calculation we obtain from (33):

$$\begin{gathered}\langle d\Sigma\rangle_{\mathbf{q}_0}=\frac{\omega_0^4}{15c^4}\left\{2|\alpha_\perp(\omega_0)-\alpha_\parallel(\omega_0)|^2+\frac{1}{2}\left[3\,|2\alpha_\perp(\omega_0)+\alpha_\parallel(\omega_0)|^2\right.\right.\\ \left.\left.+2\,|\alpha_\perp(\omega_0)|^2+|\alpha_\parallel(\omega_0)|^2\right]\sin^2\psi\right\}d\Omega.\end{gathered}\tag{37}$$

Preceding expression coincides with the result obtained for the scattering of monochromatic wave by the collection of chaotically oriented metallic nanospheroids [16].

The ultrashort electromagnetic wave is characterized by its carrier frequency $\omega_0$ and time duration $\tau$ (besides the amplitude $\mathbf{E}_0$ that doesn't influence the cross-section of elastic scattering). The metallic nanoparticle, in turn, is characterized by its volume, shape, and free electrons concentration (bearing in mind our approach). In the case of spheroidal particles, the nanoparticle's shape is defined by a relation of semiaxes (or eccentricity). Let us analyze how formula (33) depends on these parameters. To do this, we normalize the cross-section by surface area of the nanoparticle. The surface area of the spheroid (ellipsoid of revolution) is given by the formula [23]:

$$S_{ell} = 2\pi R_\perp \begin{cases} R_\perp + \frac{R_\parallel^2}{\sqrt{R_\parallel^2 - R_\perp^2}} \arcsin \frac{R_\parallel^2 - R_\perp^2}{R_\parallel}, & \text{if } R_\perp < R_\parallel \\ R_\perp + \frac{R_\parallel^2}{\sqrt{R_\perp^2 - R_\parallel^2}} \ln \frac{R_\perp + \sqrt{R_\perp^2 - R_\parallel^2}}{R_\parallel}, & \text{if } R_\perp > R_\parallel. \end{cases} \tag{38}$$

In the figures 1-3 we present the carrier frequency dependence of the normalized light-scattering cross-section for the collection of Au nanoparticles oriented chaotically. The illustrations are performed for different pulse durations (or values of parameter $\Gamma$) and for monochromatic wave that coincides limit case $\Gamma \to 0$.

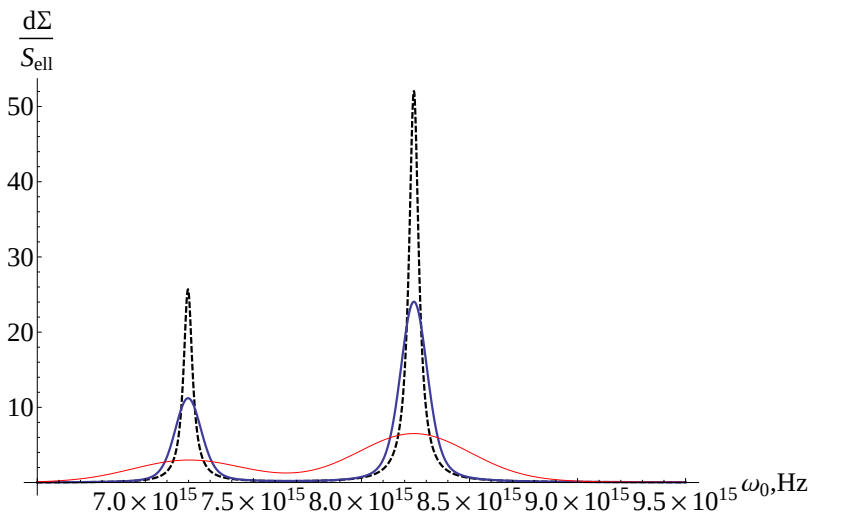

Figure 1: (Color online) The dependence of the ratio $<d\Sigma>/S_{ell}$ for prolate ($R_\perp/R_\parallel = 0.8$) spheroidal Au particles (with $R = 20\,\text{nm}$) on the carrier frequency $\omega_0$ for different values of $\Gamma$: $\Gamma = 2.5 * 10^{14}\,\text{s}^{-1} (\tau = 5\,\text{fs})$ - solid (red) line, $\Gamma = 5 * 10^{13}\,\text{s}^{-1} (\tau = 25\,\text{fs})$ - thick (blue) line, monochromatic wave ($\Gamma \to 0$) - dashed (black) line. The numerical evaluation is performed for the angle $\psi = \pi/4$.

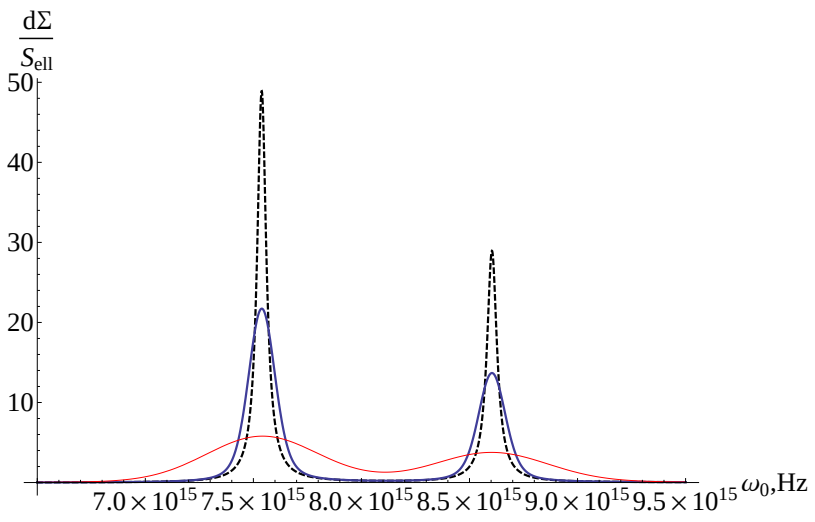

Figure 2: (Color online) The same as on Fig. 1 for oblate ($R_\perp/R_\parallel = 1.25$) particles.

The calculations were carried out using such parameters for Au: $\nu = 3.39 * 10^{13}\text{s}^{-1}$ at $0°C$ [17], $n = 5.9 * 10^{22}\text{cm}^{-3}$, $v_F = 1.39 * 10^8\text{cm/s}$ [24].

As we can see from figures 1 and 2, parameter $\Gamma$ influences the width of plasmon resonance. This result is obvious because of the meaning of $\Gamma$ as a value reciprocal to the pulse duration. As we know, the shorter the pulse is the more modes it excites. Using the inequality between energy and time, we obtain a relationship between the pulse duration $\tau$ and spectral width $\Delta\omega$:

$$\Delta\omega \cdot \tau \geq \frac{1}{2} \tag{39}$$

Equality to 1/2 in (39) can only be reached with Gaussian time and spectral envelopes. The Gaussian pulse shape consumes a minimum amount of spectral components. When the equality is reached in (39), the pulse is called a Fourier-transform-limited pulse [9]. So, we can modify formula for the half-width of a plasmon resonance. In the case of monochromatic wave, the half-width of a plasmon resonance is given by an expression:

$$\gamma_i = 2\pi L_i \sigma_{ii}(\omega). \tag{40}$$

This value strongly depends on the particles' shape and is defined by the diagonal components of the optical conductivity tensor $\sigma_{ij}$. When the laser pulse has a Gaussian shape, the half-width of a peak can be estimated as:

$$\gamma_i(\Gamma) = \Delta\omega + \pi L_i \left[\sigma_{ii}(\omega_0 - \Delta\omega) + \sigma_{ii}(\omega_0 + \Delta\omega)\right]. \tag{41}$$

The solid (red) curves in figures 1 and 2 show that two plasmonic resonances begin to join when the value of $\Gamma$ is sufficiently large ($\Gamma = 2.5 * 10^{14}\,\mathrm{s}^{-1}$ for the solid curves).

Positions of plasmonic resonances don't depend on $\Gamma$ and are defined by the particle's shape. There are two plasmonic peaks for the spheroidal particle with resonance frequencies $\omega_{\perp,\parallel} = L_{\perp,\parallel}\omega_{pl}$ (for the Au particles $\omega_{pl} \approx 1.37 * 10^{16}\,\mathrm{s}^{-1}$)[17].

As follows from figures 1 and 2, the laser pulses of longer duration are scattered more effectively. The maximum is achieved when the electromagnetic wave is monochromatic. This is a consequence of plasmonic resonance broadening – the square under the curve should be a conserved quantity. As for the shape dependence of the scattering, the maximum is achieved when $R_\perp/R_\parallel = 1$, i.e., when the particles have the spherical shape.

The figure 3 presents a surface that demonstrates dependence of the normalized scattering cross-section $<d\Sigma>/S_{ell}$ on the carrier frequency $\omega_0$ and the semiaxes ratio $R_\perp/R_\parallel$. This surface has four ranges that show how the frequencies of plasmon resonances depend on the particle's shape. These 'ranges' join in the maximum of the scattering that is observed for the ensemble of almost spherical nanoparticles at the frequency of plasmonic resonance for a sphere $\omega_{pl}/\sqrt{3}$. The point is that at the frequency $\omega_{pl}/\sqrt{3}$ the cross-section $<d\Sigma>/S_{ell}$ as a function of the semiaxes ratio $R_\perp/R_\parallel$ has a local minimum at $R_\perp/R_\parallel = 1$ and two maxima with $R_\perp/R_\parallel < 1$ and $R_\perp/R_\parallel > 1$ close to this minimum. This situation is illustrated in figure 4. Here we need to stress the point that the cross-section $<d\Sigma>/S_{ell}$ as a function of the ratio $R_\perp/R_\parallel$ demonstrates such a behavior only for ultrashort pulses, while the cross-section of a monochromatic wave has an absolute maximum at $\omega = \omega_{pl}/\sqrt{3}$ and $R_\perp/R_\parallel = 1$. The value of parameter $\Gamma = 2.5 * 10^{14}\,\mathrm{s}^{-1}$ (in figures 3 and 4) corresponds to the pulse duration $\tau \approx 5\,\mathrm{fs}$ according to the (35).

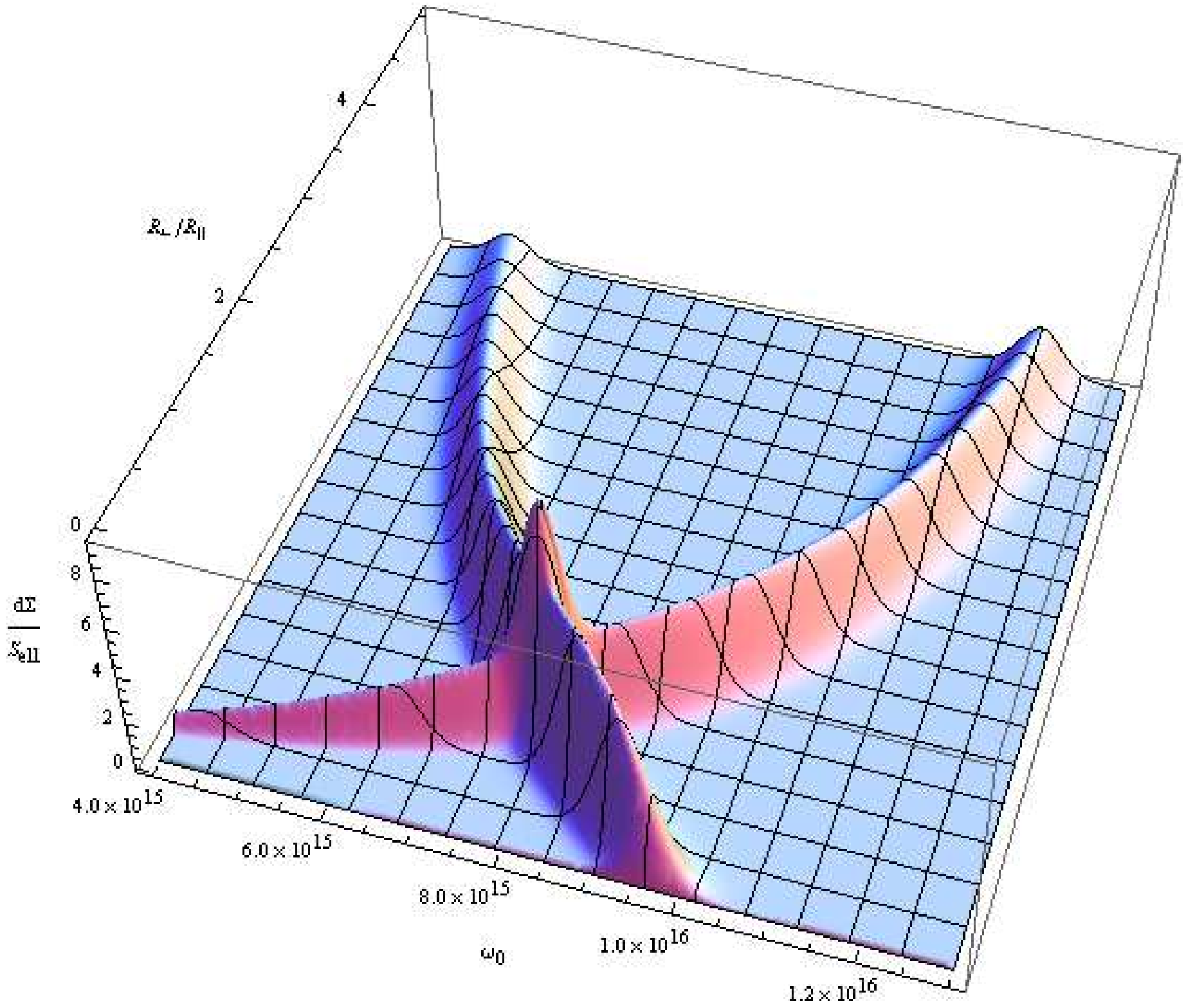

Figure 3: (Color online) The dependence of the ratio $<d\Sigma>/S_{ell}$ for spheroidal Au particles (with $R = 20\,\mathrm{nm}$) on the carrier frequency $\omega_0$ and the semiaxes ratio $R_\perp/R_\parallel$ for $\Gamma = 2.5 * 10^{14}\,\mathrm{s}^{-1}$. The numerical evaluation is performed for the angle $\psi = \pi/4$.

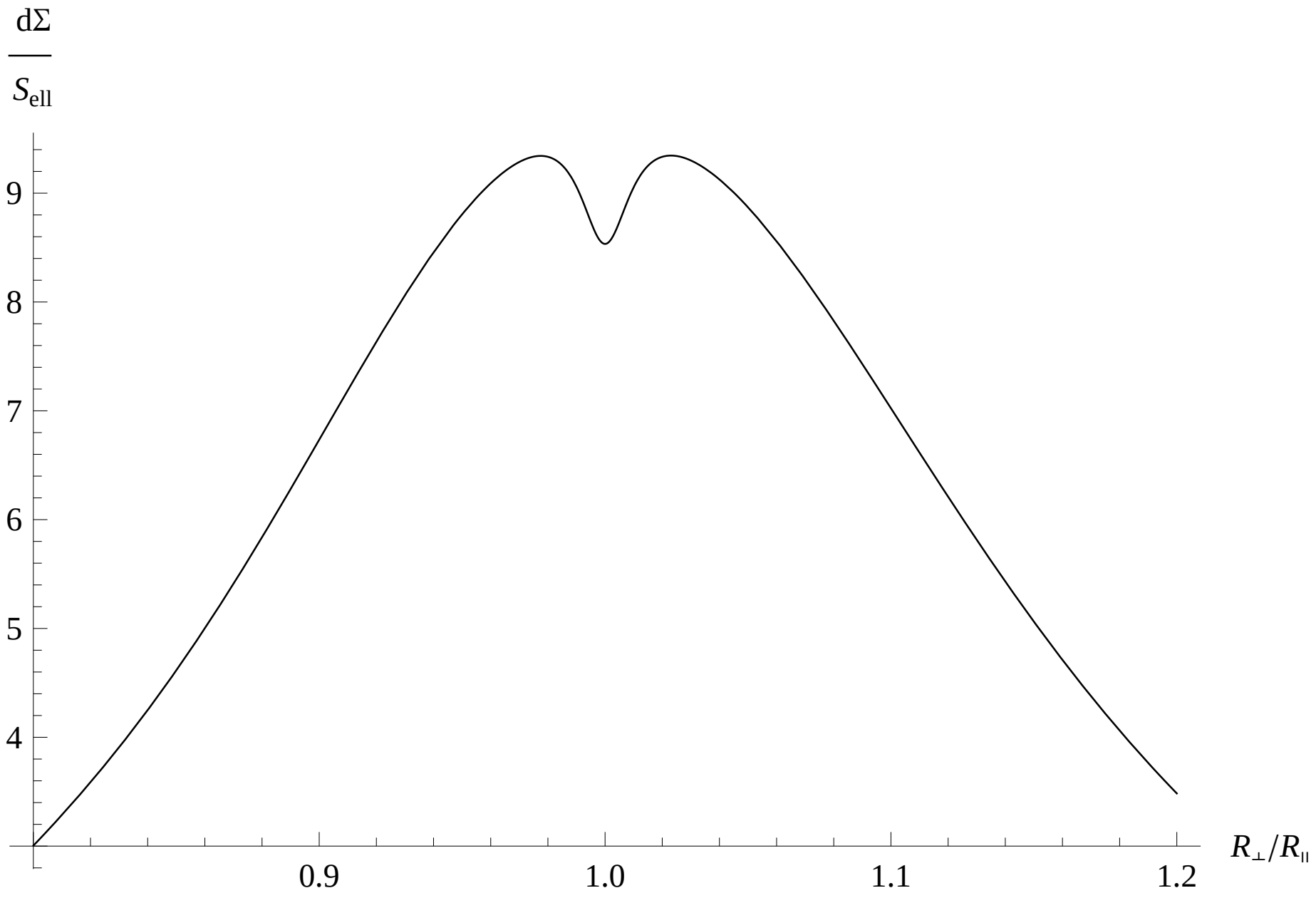

Figure 4: The dependence of the ratio $<d\Sigma>/S_{ell}$ for spheroidal Au particles (with $R = 20\,\mathrm{nm}$) on the semiaxes ratio $R_\perp/R_\parallel$ for $\Gamma = 2.5 * 10^{14}\,\mathrm{s}^{-1}$. The numerical evaluation is performed for the angle $\psi = \pi/4$.

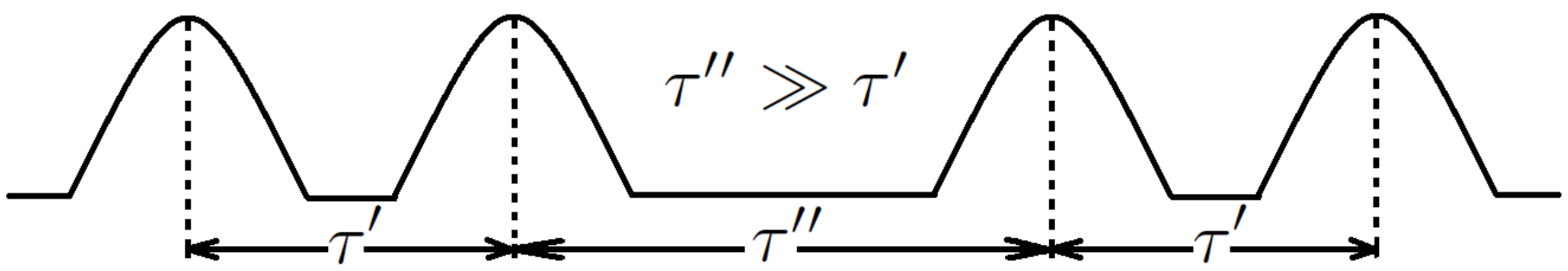

Figure 5: The double pulse scheme.

*4.2. Double ultrashort pulse*

Due to the dominant contribution to the light scattering of the collective electron motion at the wave frequencies close to the frequency of plasmon resonance, the study of nanoparticles' interaction with double ultrashort pulse becomes actual. As the first pulse of the couple excites plasmon oscillations of an electron gas, the second one can both enhance or weaken the scattering effect depending on the phase of plasmon oscillations in the moment when the second pulse is coming. As we have mentioned earlier, plasmon frequencies are explicitly defined by the nanoparticle's shape. So, when the shape is specified, the light scattering cross-section depends on the time delay between first and second pulses.

We consider a series of double ultrashort pulses with the same duration $\tau$ and time delay $\tau'$ between pulses in a couple (see figure 5). The time $\tau''$ between two couples is much bigger than time delay $\tau'$ and lifetime of plasmon oscillations. So, when the next couple of pulses acts on the nanoparticle, the nanoparticle already 'forgot' an action of previous couple by this time. An electric field of a couple is given by a superposition of single pulses' fields:

$$\mathbf{E}_T(\mathbf{r}, t) = \mathbf{E}_1(\mathbf{r}, t) + \mathbf{E}_2(\mathbf{r}, t), \tag{42}$$

where electric field of the second and first pulses are related by following expression:

$$\mathbf{E}_2(\mathbf{r}, t) = \mathbf{E}_1(\mathbf{r}, t + \tau'). \tag{43}$$

So, the Fourier component of total field can be written in such a way:

$$\mathbf{E}_T(\mathbf{r}, \omega) = \mathbf{E}_1(\mathbf{r}, \omega) \left(1 + e^{i\omega\tau'}\right). \tag{44}$$

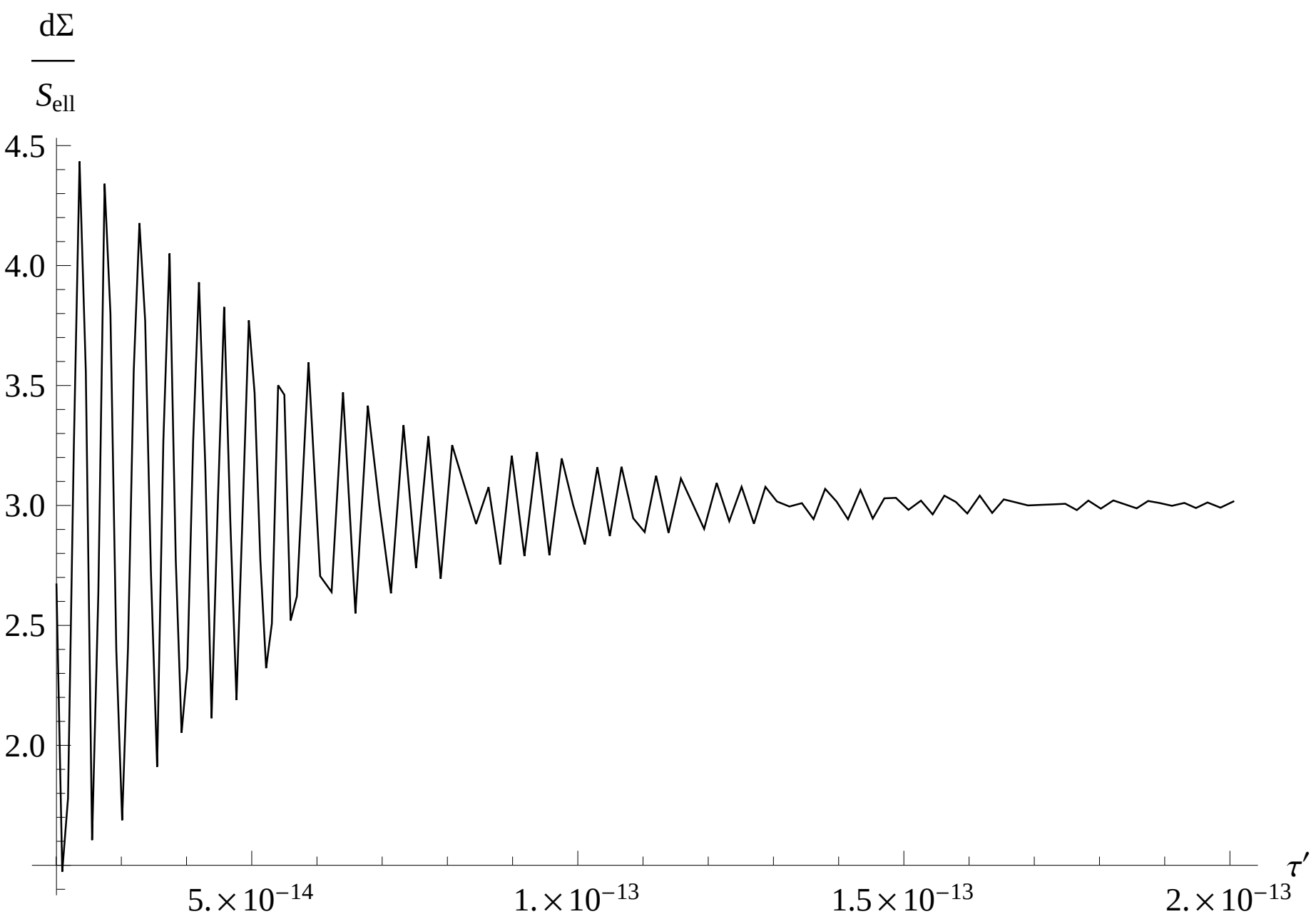

Figure 6: The dependence of the ratio $<d\Sigma> /S_{ell}$ for prolate ($R_\perp/R_\parallel = 0.8$) spheroidal Au particles (with $R = 20\,\mathrm{nm}$) on the time delay $\tau'$ for $\Gamma = 2.5 * 10^{14}\,\mathrm{s}^{-1}(\tau = 5\,\mathrm{fs})$. The numerical evaluation is performed for the angle $\psi = \pi/4$.

The quantity $\mathbf{E}_1(\mathbf{r}, \omega)$ in the last expression is given by formula (2), that's why one can obtain from (44):

$$|\mathbf{E}_T(0,\omega)|^2 = 2(1 + \cos\omega\tau')\,|\mathbf{E}(0,\omega)|^2\,. \tag{45}$$

For the double ultrashort pulse, an expression (33) for the cross-section of light scattering remains true if $|\mathbf{E}(0,\omega)|^2$ is replaced by $|\mathbf{E}_T(0,\omega)|^2$ in (31) and $\Delta I$ is calculated for the double pulse. According to (33) and (45) we obtain the energy flux density of an incident wave for the ultrashort double pulse:

$$\Delta I = \frac{c}{4\sqrt{2\pi}\Gamma}\left[1 + e^{-\omega_0^2/2\Gamma^2}\left(1 + e^{-\tau'^2\Gamma^2/2}\right) + e^{-\tau'^2\Gamma^2/2}\cdot\cos\omega_0\tau'\right]\cdot\mathbf{E}_0^2. \tag{46}$$

When the exponent $\tau'^2\Gamma^2/2 \gg 1$, or in other words $\tau' \gg \tau$, the energy flux density (46) is twice as large as the density of a single pulse (33).

The normalized cross-section dependence on the time delay $\tau'$ is illustrated in figure 6. The time range is approximately from 20 fs to 200 fs that corresponds to $4\tau$ and $40\tau$ respectively according to the formulae (35) (for $\Gamma = 2.5{*}10^{14}\,\mathrm{s}^{-1}$). The frequency of a plasmon resonance for a sphere $\omega_{pl}/\sqrt{3}$ is taken as a carrier frequency, the nanoparticles' shape is a prolate spheroid with semiaxes ratio $R_\perp/R_\parallel = 0.8$.

This figure shows that a value of the cross-section (or intensity of scattered wave) rapidly oscillates, when the time delay between pulses $\tau'$ is of the one order with pulses' duration $\tau$. These oscillations damp when the time delay $\tau'$ increases. The second pulse doesn't feel the influence of the first pulse, when the time delay is greater than the lifetime of the plasmon vibrations; and in this case, the scattering cross-section of the double ultrashort pulse turns to the same quantity for two detached ultrashort pulses at the limiting case $\tau'/\tau \gg 1$. The cross-section's oscillations damping is associated with plasmon vibrations damping. The calculations show that scattering oscillations damping doesn't depend on the pulses' duration $\tau$ and is defined by the shape of the nanoparticle. This result follows from the aforementioned fact that the frequency of plasmon vibrations doesn't depend on $\Gamma$, and $\tau$ so on.

## 5. Conclusions

In the framework of kinetic approach we study the dependence of ultrashort pulse scattered by the ensemble of metallic nanoparticles on the carrier wave frequency, pulse duration, time delay between pulses, and particles' shape. We take into account that when the size of the particle is comparable with the free electron path, one has to consider conductivity as a tensor. The product of the diagonal component of this tensor and the corresponding depolarization factor defines the half-width of the plasmonic resonance. This is the difference between our approach and Mie theory, where the conductivity is treated as a scalar and the particle's shape dependence in resonances comes only by the depolarization factors.

The calculations show that intensity of scattered wave at the frequencies of plasmonic resonances decreases when the pulse duration shortens, while the intensity of the incident wave is constant. At the same time, the half-widths of plasmonic resonances grow, and, as a result, the intensity of scattered wave increases at the frequencies that are close to the resonances. This effect is related to the fact that the shorter the pulse, the more vibration modes in the particle it excites. The positions of plasmonic resonances do not depend on the pulse duration and is defined by the nanoparticle's shape.

There is a qualitative difference between the ultrashort pulse and monochromatic wave at the frequency of plasmonic resonance for a sphere. For the monochromatic wave the maximum in scattering is observed for spherical particles, meaning that when the shape of nanoparticles deviates from spherical, intensity of the scattered field decays. For the ultrashort pulse at the same carrier frequency, two maxima in scattering are observed for prolate and oblate spheroids that are close to sphere. The local minimum, that corresponds to the spherical particle, is situated between these maxima. This effect occurs due to the presence of two plasmonic resonance in spheroidal particles. When the spheroid is close to the sphere, resonance frequencies are close too, and both of these resonances are excited by the ultrashort pulse. The total contribution to the scattering of two resonances is greater than the contribution of one plasmonic resonance that is observed for the spherical nanoparticle of the same volume.

The study of double ultrashort laser pulse's scattering by the ensemble of metallic nanoparticles shows that an intensity of scattered field oscillates with change of the time delay between the pulses. These oscillations are related to the surface plasmon oscillations in metallic nanoparticles. An intensity of the scattered light can be both enhanced or weakened depending on the plasmon oscillation's phase at the moment when the second pulse arrives. When the time delay is comparable with pulses' duration, the rapid oscillations of intensity are observed. They damp when the time delay becomes much greater than pulses' duration. So, when the time delay is small, the second pulse feels the influence of the first one via the plasmon oscillations in nanoparticles. When the time delay between pulses is greater than the plasmon lifetime, the total intensity of scattered wave is given by simple sum of two independent pulses.

## References

[1] J.H. Park, C. Park, H.S. Yu, J. Park, S. Han, J. Shin, S.H. Ko, K.T. Nam, Y.H. Cho, and Y.K. Park, Nature Photonics 95 (2013) 1.

[2] T. Nagao, G. Han, C.V. Hoang, J.-S. Wi, A. Pucci, D. Weber, F. Neubrech, V.M. Silkin, D. Ender, O. Saito, and M. Rana, Sci. Technol. Adv. Mater. 11 (2010) 054506.

[3] J. Zhao, L.J. Sherry, G.C. Schatz, and R.P. Van Duyne, IEEE J. of Selected Topics in Quantum Electronics 14 (2008) 1418.

[4] P. Alivisatos, Nat. Biotechnol. 22 (2004) 51.

[5] E. Osbay, Science 311 (2006) 5189.

[6] Y.B. Zheng, Y.-W. Yang, L. Jensen, L. Fang, B.K. Juluri, A.H. Flood, P.S. Weiss, J.F. Stoddart, and T.J. Huang, Nano Lett. 9 (2009) 819.

[7] P. Bharadway, B. Deutsh, and L. Novotny, Adv. Opt. Photonics 1 (2009) 438.

[8] N.L. Rosi and C.A. Mirkin, Chem. Rev. 105 (2005) 1547.

[9] C. Rulli'ere, Femtosecond Laser Pulses: Principles and Experiments, Springer Science+Business Media, New York, 2005.

[10] M. Kauranen and A. V. Zayats, Nature Photonics 244 (2012) 737.

[11] J.-C. Daniels and W. Rudolph, Ultrashort Laser Pulse Phenomena, Academic, New York, 1996.

[12] A.B. Shvartsburg, Time-Domain Optics of Ultrashort Wave-forms, Clarendon Press, Oxford, 1996.

[13] V.V. Temnov, Nature Photonics 220 (2012) 728.

[14] M. Sivis, M. Duwe, B. Abel, and C. Ropers, Nature Physics 2590 (2013) 1.

[15] P.M. Tomchuk and N.I. Grigorchuk, Phys. Rev. B 73 (2006) 155423.

[16] P.M. Tomchuk and D.V. Butenko, Surf. Sci. 606 (2012) 1892.

[17] N.I. Grigorchuk and P.M. Tomchuk, Phys. Rev. B 80 (2009) 155456.

[18] L.D. Landau and E.M. Lifshits, Electrodynamics of Continuous Media, Pergamon, New York, 1984.

[19] N.I. Grigorchuk and P.M. Tomchuk, Phys. Rev. B 84 (2011) 085448.

[20] A.Z. Dolginov, Yu.N. Gnedin, N.A. Silantev, Propagation and polarization of radiation in cosmic media, Gordon and Beach, Basel, 1995.

[21] M. Kerker,The scattering of light, and other electromagnetic radiation, Academic Press, New York, 1969.

[22] V. B. Berestetsky, E. M. Lifshits, L. P. Pitaevsky, Quantum Electrodynamics, Science, Moscow, 1989.

[23] G. Arfken, Mathematical Methods for Physicists, Academic Press, New York, 1986.

[24] C. Kittel, Introduction to Solid State Physics, sixth ed., Wiley, New York, 1986.

[25] N.I. Grigorchuk and P.M. Tomchuk, Low Temp. Phys. 33 (2007) 341.